\documentclass[11pt]{amsart}%
\usepackage{amssymb,latexsym,bm}
\usepackage{amsmath,color}
\usepackage{amsthm}
\usepackage{epsfig}
\usepackage{graphicx}
\usepackage{hyperref}
\usepackage{titletoc}
\usepackage{fullpage}
\usepackage{palatino,mathpazo}
\usepackage{tikz}
\usepackage{amscd}
\usepackage{multirow}
\usepackage{longtable}
\usepackage{cancel}
\usepackage{multirow}
\usepackage{enumerate}
\usepackage{bm}
\usepackage{amsmath}
\usepackage{amsfonts}
\usepackage{amssymb}%
\providecommand{\U}[1]{\protect\rule{.1in}{.1in}}
\numberwithin{equation}{section}
\newtheorem{theorem}{Theorem}[section]

\allowdisplaybreaks[4]
\usetikzlibrary{matrix}

\def\XXint#1#2#3{{\setbox0=\hbox{$#1{#2#3}{\int}$}
		\vcenter{\hbox{$#2#3$}}\kern-.5\wd0}}

\def\XXint#1#2#3{{\setbox0=\hbox{$#1{#2#3}{\int}$ }
		\vcenter{\hbox{$#2#3$ }}\kern-.6\wd0}}

\begin{document}
\title{Toda lattice and Riemann type minimal surfaces}
\author[C.F. Gui]{Changfeng Gui}
\address{Changfeng Gui, Department of Mathematics, Faculty of Science and Technology,
University of Macau, Taipa, Macao.}
\email{changfenggui@um.edu.mo}
\author[Yong Liu]{Yong Liu}
\address{Yong Liu, School of Mathematics and Statistics, Beijing Technology and
Business University, Beijing, China.}
\email{yongliu@btbu.edu.cn}
\author[J. Wang]{Jun Wang}
\address{Jun Wang, School of Mathematical Sciences, Jiangsu University, Zhenjiang,
Jiangsu, 212013, P.R. China.}
\email{wangmath2011@126.com}
\author[W. Yang]{Wen Yang}
\address{Wen Yang, Department of Mathematics, Faculty of Science and Technology,
University of Macau, Taipa, Macao.}
\email{wenyang@um.edu.mo}
\maketitle

\begin{abstract}
Toda lattice and minimal surfaces are related to each other through Allen-Cahn
equation. In view of the structure of the solutions of the Toda lattice, we
find new balancing configuration using techniques of integrable systems. This
allows us to construct new singly periodic minimal surfaces. The genus of
these minimal surfaces equals $j(j+1)/2-1$. They are natural generalization of
the Riemann minimal surfaces, which have genus zero.

\end{abstract}

\section{ Introduction}

The Toda lattice considered in this paper is the following system of
infinitely many semilinear elliptic equations:
\begin{equation}
\Delta u_{j}=e^{u_{j-1}-u_{j}}-e^{u_{j}-u_{j+1}}\text{ in }\mathbb{R}%
^{m},\text{ }j\in\mathbb{Z}\text{.} \label{Toda}%
\end{equation}
If there are only finitely many components $u_{j}$, we get the finite
nonperiodic Toda lattice(also called Toda molecule equation, \cite{Hirota}):
\begin{equation}
\begin{cases}
\Delta u_{n}=e^{u_{n-1}-u_{n}},\\
\Delta u_{j}=e^{u_{j-1}-u_{j}}-e^{u_{j}-u_{j+1}},j=2,...,n-1,\\
\Delta u_{1}=-e^{u_{1}-u_{2}}.
\end{cases}  \label{FiToda}%
\end{equation}
These Toda systems are (formally) integrable when the dimension $m=1$ or $2.$
In higher dimensions, they are believed to be non-integrable any more.

\medskip

The theory of one dimensional Toda lattice can be found in Toda\cite{Toda}.
For 1d finite Toda system, it has been studied in many papers, for instance,
\cite{Flaschka1},\cite{Flaschka2},\cite{Kons},\cite{Moser},\cite{Olsh}, just
to mention some of them.

\medskip

Our interest in the Toda lattice stems from its connection with the following
Allen-Cahn equation%
\begin{equation}
-\Delta u=u-u^{3}\text{ in }\mathbb{R}^{n}. \label{AC}%
\end{equation}
It is proved in \cite{Manuel} that one can construct a family of entire
multiple-end solutions in $\mathbb{R}^{2}$ to the Allen-Cahn equation $\left(
\ref{AC}\right)  $ using Lyapunov-Schmidt reduction. The zero level set of
these solutions is close to suitable scaling of a given solution to the $1$d
Toda system. Similar construction works for the one-soliton solution to the 1d
Toda lattice\cite{Kowal}. It is known that Allen-Cahn equation is deeply
related to the minimal surface theory, large literature exists in this
subject. One of the earliest results in this direction is its Gamma
convergence theory. A more recent result is the proof of multiplicity one
conjecture of minimal surfaces using Allen-Cahn equation\cite{CM}. On the
other hand, in \cite{K-D-W}, entire solutions of the Allen-Cahn monotone in
one direction is constructed by using the Bombieri-De Giorgi-Giusti minimal
graph in dimension 8, while in \cite{KDW2}, minimal surfaces of finite total
curvature in $\mathbb{R}^{3}$ can be used to construct solutions of Allen-Cahn
equation with finite ends. In $\mathbb{R}^{3},$ two-end solutions of the
Allen-Cahn are natural analogy of catenoids. It is known that there exists a
family of two-end solutions to the Allen-Cahn equation\cite{Gui}. The moduli
space of these solutions is diffeomorphic to $\left(  0,1\right)  .$ Near $0,$
the zero level set of the solutions is close to $2$-component Toda system.
Near $1,$ the zero level set is close to catenoid. Therefore, in a very vague
sense, Toda lattice and minimal surfaces are connected through the moduli
space of entire solutions of the Allen-Cahn equation. It is this point of view
that will be used in this paper to construct new minimal surfaces of Riemann type.

\medskip

The classical Riemann minimal surface is a singly periodic minimal surface in
$\mathbb{R}^{3}$ of genus zero foliated by circles. Roughly speaking, it can
be regarded as a desingularization of infinitely many parallel planes
connected by catenoidal necks. A characterization of Riemann's minimal surface
in terms of the topology and number of planar ends is obtained in
\cite{Lopez}(We also refer to this paper for other characterizations obtained
before 1997). It is then proved in \cite{Meeks} that if $M$ is a properly
embedded minimal surface in $\mathbb{R}^{3}$ with genus zero, and if the
symmetry group of $M$ is infinite, then $M$ is a plane, a catenoid, a helicoid
or a Riemann minimal surface. Here the assumption of genus zero is crucial.
Note that the classification of minimal surfaces has a long history and there
are many results in this direction, see for instance \cite{FM} and the
references therein. On the existence side, singly periodic minimal surfaces
with large genus and Riemann type ends have been constructed in \cite{Ca}%
,\cite{Calla},\cite{Mora},\cite{T2} using Weierstrass representation. New
examples of high genus are also constructed in \cite{Pacard} using PDE gluing
methods. It is worth pointing out that in the singly periodic case, if we
consider its quotient by the minimal period, then the Riemann minimal surface
has one planar end, while the other examples mentioned above have at least two
planar ends.

\medskip

In this paper, we will construct new singly periodic minimal surfaces of
Riemann type, with one planar end in their quotient. Our construction is
inspired the lump type solutions of the $2+1$ Toda lattice. That is, the
system $\left(  \ref{Toda}\right)  $ with $m=2.$ This system has family of
lump type solutions which have certain translational invariant properties in
the $j$ variable. These solutions are analogy to the lump type solutions of
the KP-I equation, a classical integrable system appeared in many different
mathematical and physical settings. This analogy arises from the fact that
$2+1$ Toda lattice is a discrete version of the KP-I equation. Our recent
result \cite{Liu} shows that lump type solutions can be classified in terms of
the degree of their tau functions. These degrees have to be of the form
$j\left(  j+1\right)  .$ In particular, the tau function of the classical lump
solution has degree $2$. In view of the moduli space theory of Allen-Cahn
equation, the natural analogy of the classical lump solution will be the
Riemann minimal surfaces, while lump type solutions correspond to the singly
periodic Riemann type minimal surfaces described in the following

\begin{theorem}
\label{main}For each $R$ large enough, there exists a singly periodic minimal
surfaces with infinitely many planar ends(Riemann type end, the distance
between consecutive planes is normalized to be $1$), with period $\left(
R,0,1\right)  .$ It is obtained by desingularizing catenoids between parallel
planes, whose location is determined by the polynomial $P_{j}$ obtained in
Section 3. The degree of $P_{j}$ is equal to $\frac{j\left(  j+1\right)  }%
{2}.$
\end{theorem}

The construction uses catenoids as basic blocks, to connect parallel planes in
a periodic manner. Between two consecutive planes, we insert $\frac{j\left(
j+1\right)  }{2}$ catenoids with the same size. The resulted minimal surface
has genus $\frac{j\left(  j+1\right)  }{2}-1.$ This type of technique is by
now commonly used. The main new step of our proof is to find new balancing
configurations of these catenoids. This is done by using ideas(Backlund
transformation) from the theory of integrable systems. We can actually
construct other minimal surfaces of \textquotedblleft soliton type". Details
can be found in Section 3. The classification of these type minimal surfaces
is definitely a more difficult problem.

\medskip

In Section 2, we have collected some known facts of Allen-Cahn equation, Toda
lattices, and minimal surfaces, and also discussed the approximate minimal
surfaces we would like to construct.

{\center{\bf Acknowledgement}.} C.F. Gui is partially supported by University
of Macau research grants CPG2023-00037-FST, SRG2023-00011-FST and an UMDF
Professorial Fellowship. Y. Liu is supported by the National Key R\&D Program
of China 2022YFA1005400. J. Wang is partially supported by National Key R$\&$D
Program of China 2022YFA1005601 and NSFC No.12371114. W. Yang is partially
supported by National Key R\&D Program of China 2022YFA1006800, NSFC No.
12171456, NSFC No. 12271369 and SRG2023-00067-FST.

\bigskip

\section{Preliminaries}

Toda lattice is integrable. Its inversing scattering theory has been studied
in \cite{V1,V2,V3}. Formally, it is a discretization of the KP-I
equation\cite{Popowicz}: They are connected through a family of integrable
systems of the form
\begin{align}
&  \left\{  \left[  \partial_{t}+\left(  h^{-2}-h^{2}\right)  \partial
_{x}\right]  ^{2}+h^{-2}\partial_{y}^{2}\right\}  \ln\left(  1+h^{2}u\right)
\nonumber\\
&  =h^{-4}\left[  u\left(  x+h,y,t\right)  +u\left(  x-h,y,t\right)
-2u\right]  . \label{dis}%
\end{align}
Here $h$ is a parameter. As $h\rightarrow0,$ formally $\left(  \ref{dis}%
\right)  $ tends to the KP-I equation%
\[
\partial_{x}\left(  -\partial_{t}u+\frac{1}{2}u\partial_{x}u+\frac{1}%
{24}\partial_{x}^{3}u\right)  -\frac{1}{2}\partial_{y}^{2}u=0.
\]
We also point out that the relation between $1+1$ Toda lattice and KdV
equation has been rigorously established in \cite{Bambusi}.

In our recent paper \cite{Liu}, it is proved that if $u$ is a lump type
solution of the KP equation(a traveling wave solution decaying to zero at
space infinity), the $u=2\partial_{x}^{2}\ln f,$ where $f$ is a polynomial of
degree $n\left(  n+1\right)  .$ If in addition $u$ is even, then for fixed
$n,$ $f$ is unique. In view of the above correspondence, similar
results(existence and classification) are expected to be true for the Toda
lattice. In the $2+1$ elliptic Toda lattice $\left(  \ref{Toda}\right)  $, if
we require that
\[
u_{j}\left(  x,y\right)  =u_{j-1}\left(  x-a,y\right)  ,
\]
then it reduces to a differential-difference equation:
\[
\Delta u_{j}=\exp\left[  u_{j}\left(  x+a,y\right)  -u_{j}\left(  x,y\right)
\right]  -\exp\left[  u_{j}\left(  x,y\right)  -u_{j}\left(  x-a,y\right)
\right]  .
\]
This type of solutions correspond to the traveling wave solutions of the KP
equation. By a direct computation using the Hirota bilinear form the Toda
lattice, we have the following one parameter family of solutions of Toda
lattice with
\[
u_{j+1}\left(  x,y\right)  =u_{j}\left(  x-a,y\right)  ,
\]
of the form
\[
u_{j}\left(  x,y\right)  =\ln\frac{f\left(  x-a,y\right)  }{f\left(
x,y\right)  },
\]
where $a=\sqrt{2-\frac{2}{t}},$ $f$ is the following degree $6$ even
polynomial:
\[
b_{0}x^{6}+y^{6}+b_{1}x^{4}y^{2}+b_{2}x^{2}y^{4}+b_{3}x^{4}+b_{4}y^{4}%
+b_{5}x^{2}y^{2}+b_{6}x^{2}+b_{7}y^{2}+b_{8}.
\]
Here the coefficients $b_{j}$ are explicitly given by%
\begin{align*}
b_{0}  &  =t^{3},b_{1}=3t^{2},b_{2}=3t,b_{3}=\frac{t^{2}}{6}\left(
25t^{2}-58t+33\right)  ,\\
b_{4}  &  =\frac{1}{6}\left(  17t^{2}-26t+9\right)  ,b_{5}=15t\left(
t-1\right)  ^{2},b_{6}=-\frac{1}{36}t\left(  t-1\right)  ^{2}\left(
125t^{2}+30t-171\right)  ,\\
b_{7}  &  =\frac{1}{36}\left(  t-1\right)  ^{2}\left(  475t^{2}%
-702t+243\right)  ,b_{8}=\frac{1}{72}\left(  t-1\right)  ^{4}\left(
25t-9\right)  ^{2}.
\end{align*}
This correspond to the lump type solution of the KP-I equation whose tau
function is a degree $6$ even polynomial.

Having understood the relation between the Toda lattice and KP equation, we
proceed to discuss its relation with the minimal surfaces. The theory of 2D
Toda lattice and minimal surfaces are formally connected through the
Allen-Cahn equation%
\[
-\Delta u=u-u^{3}.
\]
Consider solutions $u$ of the Allen-Cahn equation satisfying the following
periodic condition
\[
u\left(  x,y,z\right)  =u\left(  x-e_{1},y,z-e_{3}\right)  ,\text{ for some
}e_{1},e_{3}.
\]
Moreover, we assume that the zero level set of $u$ are asymptotic to
infinitely many horizontal planes, with distance $e_{3}$ between two adjacent
planes. The space of all such solutions(moduli translation and rotation) is of
dimension two, provided that $u$ is nondegenerated in suitable sense. In the
simplest case, near the boundary of this moduli space, we should see lump
solution of Toda lattice on one end, and Riemann's minimal surface on the
other end. This inspired us to construct new minimal surfaces from the known
structure of the Toda lattice. Note that Toda lattice is an integrable system
and actually we have many tools available to investigate it. It turns out that
we can construct new minimal surfaces using this point of view. One of the
basic ingredients of our construction will be catenoid, a classical minimal
surface with two ends. Let us recall some of its properties most relevant to
our later purpose.

The catenoid is an axially symmetric minimal surface given by
\[
\left\vert z\right\vert =\ln\left(  r+\sqrt{r^{2}-1}\right)  .
\]
Let $\left(  x_{j},y_{j}\right)  ,j=1,...,m$ be $n$ distinct points, far away
from each other. We have in mind to put catenoids centered at $\left(
x_{j},y_{j}\right)  .$ Define
\[
r_{j}=\sqrt{\left(  x-x_{j}\right)  ^{2}+\left(  y-y_{j}\right)  ^{2}},
\]
and let
\[
f_{j}\left(  x,y\right)  =\sigma_{j}\ln\left(  r_{j}+\sqrt{r_{j}^{2}%
-1}\right)  .
\]
where $\left\vert \sigma_{j}\right\vert =1.$ Away from the center of the
catenoid, the minimal surfaces we would like to construct will be close to the
graph of the function
\[
f\left(  x,y\right)  :=\sum\limits_{j=1}^{m}f_{j}\left(  x,y\right)  .
\]
It is important to understand the error of $f$ as an approximate minimal
surface. Suppose $\left(  x_{1},y_{1}\right)  =\left(  0,0\right)  .$ We
compute the error of $f$ in the region $\Omega$ where $2<r<M_{0},$ where
$M_{0}$ is a fixed large constant. We have
\begin{align*}
&  \operatorname{div}\left(  \frac{\nabla\left(  f_{1}+f_{2}\right)  }%
{\sqrt{1+\left\vert \nabla\left(  f_{1}+f_{2}\right)  \right\vert ^{2}}%
}\right) \\
&  =\operatorname{div}\left(  \frac{\nabla f_{1}}{\sqrt{1+\left\vert \nabla
f_{1}\right\vert ^{2}+2\nabla f_{1}\cdot\nabla f_{2}+\left\vert \nabla
f_{2}\right\vert ^{2}}}\right) \\
&\quad  +\operatorname{div}\left(  \frac{\nabla f_{2}}{\sqrt{1+\left\vert \nabla
f_{1}\right\vert ^{2}+2\nabla f_{1}\cdot\nabla f_{2}+\left\vert \nabla
f_{2}\right\vert ^{2}}}\right)  .
\end{align*}
To simplify the notation, set
\[
\alpha=\frac{2\nabla f_{1}\cdot\nabla f_{2}+\left\vert \nabla f_{2}\right\vert
^{2}}{1+\left\vert \nabla f_{1}\right\vert ^{2}}.
\]
We first analyze%
\begin{align*}
I_{1}  &  :=\operatorname{div}\left(  \frac{\nabla f_{1}}{\sqrt{1+\left\vert
\nabla f_{1}\right\vert ^{2}+2\nabla f_{1}\cdot\nabla f_{2}+\left\vert \nabla
f_{2}\right\vert ^{2}}}\right) \\
&  =\operatorname{div}\left(  \frac{\nabla f_{1}}{\sqrt{1+\left\vert \nabla
f_{1}\right\vert ^{2}}\sqrt{1+\alpha}}\right)  .
\end{align*}
Using the fact that $f_{1},f_{2}$ satisfies the minimal surface equation, we
obtain%
\[
I_{1}=\frac{1}{\sqrt{1+\left\vert \nabla f_{1}\right\vert ^{2}}}\nabla
f_{1}\cdot\nabla\left(  \frac{1}{\sqrt{1+\alpha}}\right)  .
\]
Hence in $\Omega,$ using the fact that $\nabla f_{2}$ is small, we find that
the main order term of $I_{1}$ is
\[
-\frac{1}{2\sqrt{1+\left\vert \nabla f_{1}\right\vert ^{2}}}\nabla f_{1}%
\cdot\nabla\left(  \frac{2\nabla f_{1}\cdot\nabla f_{2}}{1+\left\vert \nabla
f_{1}\right\vert ^{2}}\right)  .
\]
On the other hand,
\begin{align*}
I_{2}  &  =\operatorname{div}\left(  \frac{\nabla f_{2}}{\sqrt{1+\left\vert
\nabla f_{1}\right\vert ^{2}+2\nabla f_{1}\cdot\nabla f_{2}+\left\vert \nabla
f_{2}\right\vert ^{2}}}\right) \\
&  =\operatorname{div}\left(  \frac{\nabla f_{2}}{\sqrt{1+\left\vert \nabla
f_{1}\right\vert ^{2}}\sqrt{1+\alpha}}\right) \\
&  =\frac{\Delta f_{2}}{\sqrt{1+\left\vert \nabla f_{1}\right\vert ^{2}}%
\sqrt{1+\alpha}}+\nabla f_{2}\cdot\nabla\frac{1}{\sqrt{1+\left\vert \nabla
f_{1}\right\vert ^{2}}\sqrt{1+\alpha}}.
\end{align*}
This implies that the main order term of $I_{2}$ in $\Omega$ will be%
\[
\frac{\Delta f_{2}}{\sqrt{1+\left\vert \nabla f_{1}\right\vert ^{2}}}+\nabla
f_{2}\cdot\nabla\frac{1}{\sqrt{1+\left\vert \nabla f_{1}\right\vert ^{2}}}.
\]
Let $\rho$ be a cutoff function such that
\[
\rho\left(  r\right)  =\left\{
\begin{array}
[c]{l}%
1,0\leq r<1,\\
0,r\geq2.
\end{array}
\right.
\]
In reality, in the region where $0<r<1,$ the approximate minimal surface $S$
will be the graph of
\[
f_{1}\left(  x,y\right)  +\sum\limits_{j=2}^{m}f_{j}\left(  0,0\right)  .
\]
For \thinspace$1\leq r\leq2,$ $S$ will be defined to be
\[
\rho\left(  r\right)  \left[  f_{1}\left(  x,y\right)  +\sum\limits_{j=2}%
^{m}f_{j}\left(  0,0\right)  \right]  +\left(  1-\rho\left(  r\right)
\right)  f\left(  x,y\right)  .
\]
Note that in $\Omega,$
\begin{align*}
\nabla f_{2}  &  =\frac{1}{r_{2}^{2}\sqrt{r^2_2-1}}\left(  x-x_{2},y-y_{2}\right)  ,\\
\Delta f_{2}  &  =\partial_{r_{2}}^{2}f_{2}+\frac{1}{r_{2}}\partial_{r_{2}%
}f_{2}=O\left(  r_{2}^{-4}\right)  .
\end{align*}
Consider the two Jacobi fields of the catenoid $f_{1}$ corresponding to the
translation along $x$ and $y$ direction. Denote them by $J_{1}$ and $J_{2}.$
Let $P_{1}$ and $P_{2}$ be the error $E\left(  S\right)  $ around $f_{1},$
onto $J_{1}$ and $J_{2}.$ Then using the previous computation and the symmetry
of $J_{1},J_{2},$ we see that at the main order
\[
P_{1}-iP_{2}=\frac{c_{0}}{x_{2}+iy_{2}},
\]
where $c_{0}$ is a fixed constant and $i$ is the imaginary unit. This is the
force of catenoid $f_{2}$ exerted onto $f_{1}.$ The balancing condition
requires that all the forces of other catenoids to $f_{1}$ to be zero. This
condition is already derived in \cite{Mora,T1,T2} in the framework of
Weierstrass representation, and used there to construct many new minimal
surfaces by desingularize a balanced configuration. For this reason, although
here we are focused on a PDE point of view, we will not delve into more
details, and refer the interested readers to \cite{Pacard,Mora,T1,T2}. The
advantage of PDE method is that it is possible to extend the results to higher
dimensions, as was done in \cite{Pa2}.

\section{Balanced configuration and existence of minimal surfaces}

In this section, we would like to construct singly periodic minimal surfaces
of Riemann type by using some special configuration of parallel planes
connected by certain number of catenoids.

The coordinate of $\mathbb{R}^{3}$ will be denoted by $\left(  x_{1}%
,x_{2},x_{3}\right)  .$ Assume that we have infinitely many horizontal planes
$L_{k},$ $k\in\mathbb{Z},$ explicitly described by $x_{3}=k.$

For $j=1,...,n,$ let us use complex numbers $p_{j}$ to denote the location of
$n$ catenoids between $L_{0}$ and $L_{1}$. These points $p_{j}$ will be
assumed to be distinct.

We introduce the notation
\[
p_{j}^{+}=p_{j}+1,\text{ \ }p_{j}^{-}=p_{j}-1.
\]
The points $p_{j}^{+}$ will represent the location of the catenoinds between
$L_{1}$ and $L_{2};$ while $p_{j}^{-}$ will be the location of the catenoinds
between $L_{-1}$ and $L_{0}.$ Between the planes $L_{k}$ and $L_{k+1},$ the
location of catenoids will be $p_{j}+k.$

Balancing condition requires that for each fixed index $k\in\left\{
1,...,n\right\}  ,$%
\begin{equation}
\sum\limits_{j\neq k}\frac{2}{p_{k}-p_{j}}=\sum\limits_{j=1}^{n}\left(
\frac{1}{p_{k}-p_{j}^{+}}+\frac{1}{p_{k}-p_{j}^{-}}\right)  . \label{ba}%
\end{equation}
This is system of algebraic equations. For $n>1,$ it is not easy to be
directly solved. In particular, it is not clear that for which $n,$ it has a
solutions. To investigate this problem, we define the generating functions
\begin{align*}
P\left(  z\right)   &  =\prod\limits_{j=1}^{n}\left(  z-p_{j}\right)  ,\\
P^{+}\left(  z\right)   &  =\prod\limits_{j=1}^{n}\left(  z-p_{j}^{+}\right)
,\\
P^{-}\left(  z\right)   &  =\prod\limits_{j=1}^{n}\left(  z-p_{j}^{-}\right)
.
\end{align*}
Direct computation tells us
\begin{equation}
P^{\prime}\left(  z\right)  =P\left(  z\right)  \sum\limits_{j=1}^{n}\frac
{1}{z-p_{j}},\text{ } \label{Pri}%
\end{equation}
and a further differentiation yields
\begin{align*}
\text{\ }P^{\prime\prime}\left(  z\right)   &  =P\left(  z\right)
\sum\limits_{j,k,k\neq j}\left(  \frac{1}{z-p_{k}}\frac{1}{z-p_{j}}\right) \\
&  =-P\left(  z\right)  \sum\limits_{k=1}^{n}\left(  \frac{1}{z-p_{k}}
\sum\limits_{j,j\neq k}\frac{2}{p_{j}-p_{k}}\right)  .
\end{align*}
In view of the balancing condition $\left(  \ref{ba}\right)  ,$ we have
\[
\frac{P^{\prime\prime}}{P}=-\sum\limits_{k=1}^{n}\left[  \frac{1}{z-p_{k}}%
\sum\limits_{j=1}^{n}\left(  \frac{1}{p_{j}^{+}-p_{k}}+\frac{1}{p_{j}
^{-}-p_{k}}\right)  \right]  .
\]
On the other hand, using $\left(  \ref{Pri}\right)  $ we obtain
\begin{align*}
\frac{P^{\prime}}{P}\frac{P^{+\prime}}{P^{+}}  &  =\sum\limits_{k=1}^{n}
\frac{1}{z-p_{k}}\sum\limits_{j=1}^{n}\frac{1}{z-p_{j}^{+}}\\
&  =-\sum\limits_{k=1}^{n}\frac{1}{z-p_{k}}\sum\limits_{j=1}^{n}\frac{1}
{p_{j}^{+}-p_{k}}+\sum\limits_{k=1}^{n}\frac{1}{z-p_{k}^{+}}\sum
\limits_{j=1}^{n}\frac{1}{p_{j}-p_{k}^{+}}.
\end{align*}

Introduce the notation
\[
\sum\limits^{\ast}f\left(  x\right)  =\sum\limits_{j=-\infty}^{+\infty
}f\left(  x+j\right)  .
\]
We arrive at
\begin{equation}
\sum\limits^{\ast}\frac{P^{\prime\prime}}{P}-\sum\limits^{\ast}\left(
\frac{P^{\prime}}{P}\frac{P^{+\prime}}{P^{+}}\right)  =0. \label{infisum}%
\end{equation}
Thus
\[
\sum\limits^{\ast}\frac{P^{\prime\prime}P^{+}-P^{\prime}P^{+\prime}}{PP^{+}%
}=0.
\]
We are lead to find polynomial $P$ and $\bar{Q}$ such that%

\begin{equation}
P^{\prime\prime}P^{+}-P^{\prime}P^{+\prime}+P\bar{Q}^{+}-P^{+}\bar{Q}=0.
\label{polyeq}%
\end{equation}

To get a grasp on this equation, let us try the method of undetermined
coefficients. Let us assume the degree of $P$ is equal to $n,$ and the degree
of $\bar{Q}$ is equal to $m.$ Then highest degree term in $P^{\prime\prime
}P^{+}-P^{\prime}P^{+\prime}$ is
\[
\left(  n\left(  n-1\right)  -n^{2}\right)  z^{2n-2}=-nz^{2n-2}.
\]
Note that it never vanishes. On the other hand, observe that the highest
degree term in $\bar{Q}P^{+}-\bar{Q}^{+}P$ is $z^{m+n-1},$ rather than
$z^{m+n}.$ Therefore, $m$ should satisfies
\[
m+n-1=2n-2.
\]
This tells us that the degree $m$ of $\bar{Q}$ should satisfy $m=n-1.$

In the case of $n=3,$ we find the following solutions
\begin{align*}
P\left(  z\right)   &  =z^{3}+az^{2}+\frac{a^{2}-1}{3}z.\\
\bar{Q}\left(  z\right)   &  =3z^{2}+\left(  2a-3\right)  z+\frac{1}{3}
(a^{2}-3a-4),
\end{align*}
where $a$ is a complex parameter. The dimension of the configuration space is
thus equal to $4.$

Note that $z=0$ is a root of $P,$ and
\[
P\left(  z\right)  =z\left(  z^{2}+az+\frac{a^{2}-1}{3}\right)  ,
\]
We can take $a=0$ and get a symmetric configuration, corresponds to the roots
of equation
\[
z^{2}-\frac{1}{3}=0.
\]
That is $z=\pm\frac{\sqrt{3}}{3}.$ In this case,
\[
\bar{Q}\left(  z\right)  =3z^{2}-3z-\frac{4}{3}.
\]

In the case of $n=4$, we find that no solution exists. Indeed,
\[
P\left(  z\right)  =z^{4}+2z^{3}+z^{2}=z^{2}\left(  z+1\right)  ^{2}.
\]

Next let us consider the case of $n=6.$ In this case, we find the following
solution
\begin{align*}
P\left(  z\right)   =~&z^{6}+\left(  a+6t\right)  z^{5}+\left(  -\frac{5}
{3}+\frac{5a^{2}}{12}+5at+15t^{2}\right)  z^{4}\\
&  +A_{3}z^{3}+A_{2}z^{2}+A_{1}z+A_{0},
\end{align*}
where
\begin{align*}
A_{3}    =&-\frac{10a}{9}+\frac{5a^{3}}{72}-\frac{1}{36}\sqrt{64-80a^{2}
+25a^{4}+1080ae}\\
&  -\frac{20t}{3}+\frac{5a^{2}t}{3}+10at^{2}+20t^{3}.
\end{align*}
$A_{2},A_{1},A_{0}$ are also depending on $a,t,e.$ We are looking for a
symmetric configuration. This requires $a=-6t$
\[
e=\frac{\left(  2-45t^{2}\right)  ^{2}-\left(  45t^{3}\right)  ^{2}}{405t},
\]%
\[
t^{6}-\frac{5}{3}t^{4}+\frac{4}{9}t^{2}+\frac{4}{405}=0.
\]
Then
\[
P\left(  z\right)  =z^{6}-\frac{5}{3}z^{4}+\frac{4}{9}z^{2}+\frac{4}{405}.
\]
Solutions of this equation is
\[
\pm1.151,\pm0.6012,\pm0.1435i.
\]
The corresponding polynomial $\bar{Q}$ is
\[
6z^{5}-15z^{4}-\frac{35}{3}z^{3}+10z^{2}+\frac{14}{9}z-\frac{4}{9}.
\]

\begin{figure}[ptb]
\centering
{ \includegraphics[
		height=0.5 \textheight,
		width=0.5\textheight
			\caption{The roots of $P$ with degree $6$.}
		]{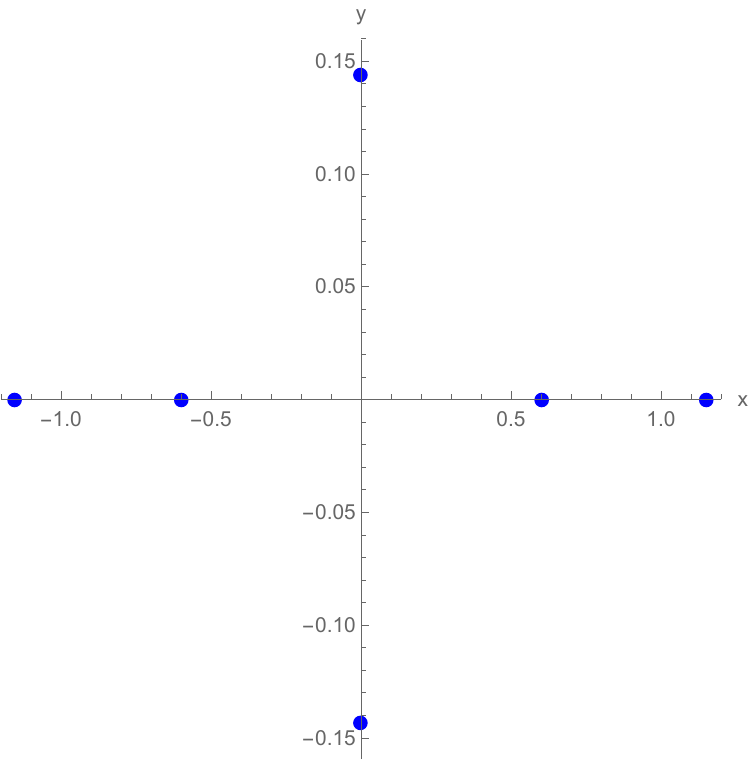} }\end{figure}

\begin{figure}[ptb]
\centering
{ \includegraphics[
		height=0.5 \textheight,
		width=0.5\textheight
			\caption{The roots of $P, P^{+},P^{-}$ with degree $6$.}
		]{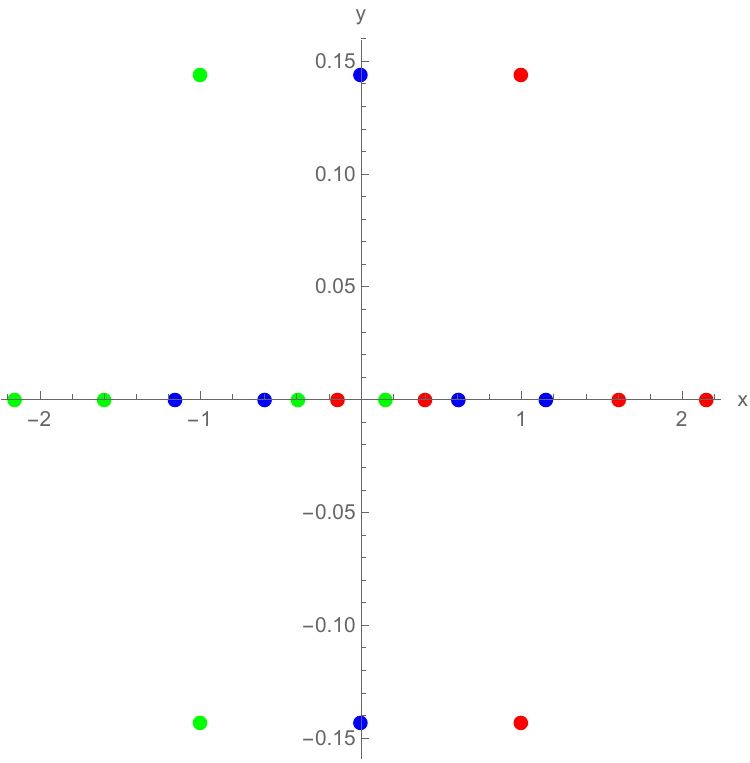} }\end{figure}

\begin{figure}[ptb]
\centering
{ \includegraphics[
		height=0.5  \textheight,
		width=0.5\textheight
		\caption{The roots of $P$ with degree $10$.}
		]{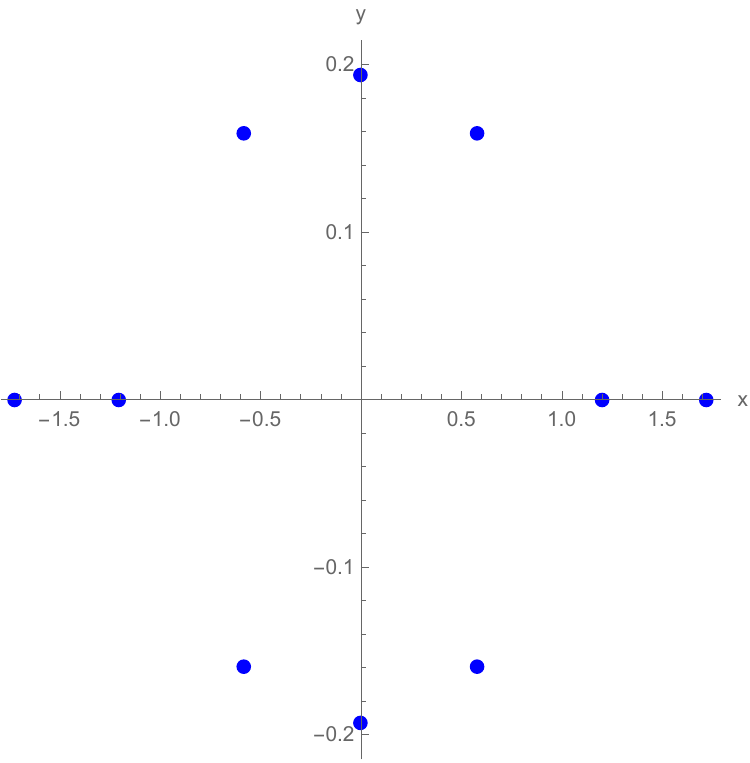} }\end{figure}Note that both the polynomials $P$ and $\bar
{Q}$ have rational coefficients. This is expected to be true of all $n.$

It is conjectured that a solution exists if and only if $n=\frac{k\left(
k+1\right)  }{2},$ where $k$ is a positive integer. That is, when $n$ is a
triangular number.

In the case of $n=10,$ we find the unique polynomial solution(without repeated
roots)
\[
P\left(  z\right)  =z^{10}-5z^{8}+7z^{6}-3z^{4}+\frac{4}{9}z^{2}+\frac{4}%
{189}.
\]
In this case,
\[
\bar{Q}\left(  z\right)  =10z^{9}-45z^{8}-55z^{7}+140z^{6}+70z^{5}%
-105z^{4}-27z^{3}+18z^{2}+\frac{38}{9}z-\frac{4}{9}.
\]
Note that actually there are other polynomial solutions with repeated roots.

Unfortunately, the above method does not work when $n=15$. To solve the
balancing system for large $n,$ we need to use different method. Observe that
\[
\sum\limits^{\ast}\frac{P^{\prime\prime}}{P}-\sum\limits^{\ast}\left(
\frac{P^{\prime}}{P}\frac{P^{-\prime}}{P^{-}}\right)  =0.
\]
Replacing $z$ be $z-1,$ we get
\[
\sum\limits^{\ast}\frac{P^{+\prime\prime}}{P^{+}}-\sum\limits^{\ast}\left(
\frac{P^{+\prime}}{P^{+}}\frac{P^{\prime}}{P}\right)  =0.
\]
Hence
\[
\sum\limits^{\ast}\frac{PP^{+\prime\prime}-P^{\prime}P^{+\prime}}{PP^{+}}=0.
\]
This implies the existence of another polynomial $\tilde{Q}$ such that
\begin{equation}
PP^{+\prime\prime}-P^{\prime}P^{+\prime}+P\tilde{Q}^{+}-P^{+}\tilde{Q}=0.
\label{polyeq2}%
\end{equation}
Combining $\left(  \ref{polyeq}\right)  $ and $\left(  \ref{polyeq2}\right)
$, we get
\begin{equation}
P^{\prime\prime}P^{+}-2P^{\prime}P^{+\prime}+PP^{+\prime\prime}+PQ^{+}
-P^{+}Q=0, \label{PQ}%
\end{equation}
where $Q=\bar{Q}+\tilde{Q}.$ The advantage of this equation is that it can be
written in the Hirota form.

Let $D$ be the Hirota bilinear derivative operator. The equation \eqref{PQ}
can be written in the form
\begin{equation}
D^{2}P\cdot P^{+}+PQ^{+}-P^{+}Q=0. \label{bili}%
\end{equation}
We shall solve this equation using Backlund transformation. Suppose $\left(
f,g\right)  $ and $\left(  F,G\right)  $ both satisfy $\left(  \ref{bili}%
\right)  :$
\begin{equation}
D^{2}f\cdot f^{+}+fg^{+}-f^{+}g=0. \label{f}%
\end{equation}%
\begin{equation}
D^{2}F\cdot F^{+}+FG^{+}-F^{+}G=0. \label{F}%
\end{equation}
Using equation (I.4) of \cite{Rogers}, we obtain
\begin{align*}
&  \left[  D^{2}f\cdot f^{+}+fg^{+}-f^{+}g\right]  FF^{+}-\left[  D^{2}F\cdot
F^{+}+FG^{+}-F^{+}G\right]  ff^{+}\\
&  =\left(  D^{2}f\cdot f^{+}\right)  FF^{+}-\left(  D^{2}F^{+}\cdot F\right)
ff^{+}\\
& \quad +\left(  fg^{+}-f^{+}g\right)  FF^{+}-\left(  FG^{+}-F^{+}G\right)
ff^{+}\\
&  =D\left[  \left(  Df\cdot F\right)  \cdot f^{+}F^{+}-fF\cdot\left(
Df^{+}\cdot F^{+}\right)  \right] \\
& \quad +\left(  fg^{+}-f^{+}g\right)  FF^{+}-\left(  FG^{+}-F^{+}G\right)  ff^{+}.
\end{align*}
Let us write
\begin{align*}
&  \left(  fg^{+}-f^{+}g\right)  FF^{+}-\left(  FG^{+}-F^{+}G\right)  ff^{+}\\
&  =\left(  g^{+}F^{+}-G^{+}f^{+}\right)  fF-\left(  gF-Gf\right)  f^{+}F^{+}.
\end{align*}
Note that $f,F$ satisfies $\left(  \ref{f}\right)  $ and $\left(
\ref{F}\right)  $ is equivalent to
\begin{align*}
&  D\left[  \left(  Df\cdot F\right)  \cdot f^{+}F^{+}-fF\cdot\left(
Df^{+}\cdot F^{+}\right)  \right] \\
&  =\left(  g^{+}F^{+}-G^{+}f^{+}\right)  fF-\left(  gF-Gf\right)  f^{+}F^{+}.
\end{align*}
Introducing two new functions $\phi$ and $\Phi,$ we look for solutions
satisfies one of the following systems:
\begin{equation}
\begin{cases}
f=\phi^{+},F=-\Phi,\\
Df\cdot F=\phi^{+}\Phi-\phi\Phi^{+},\\
gF-Gf=-D\phi^{+}\cdot\Phi-D\phi\cdot\Phi^{+}.
\end{cases}.  \label{s1}%
\end{equation}
or
\begin{equation}
\begin{cases}
f=-\phi,F=\Phi^{+},\\
Df\cdot F=\phi^{+}\Phi-\phi\Phi^{+},\\
gF-Gf=-D\phi^{+}\cdot\Phi-D\phi\cdot\Phi^{+}.
\end{cases}  \label{s2}%
\end{equation}
It is not clear to us at this moment what is the relation between our Backlund
transformation and that of the Toda lattice discussed in \cite{HS}.

In the case of $f\left(  z\right)  =z+c_{1}:=P_{1}\left(  z\right)  $ and
$g\left(  z\right)  =2,$ they satisfies $\left(  \ref{f}\right)  .$ We use
$\left(  \ref{s1}\right)  .$ Let $f=\phi^{+},F=-\Phi,$ we can solve the
equation%
\[
Df\cdot F=-fF+f^{-}F^{+}.
\]
This is a first order differential-difference linear equation for the unknown
polynomial function $F.$ We then obtain the solution
\[
F\left(  z\right)  =z^{3}+\left(  3c_{1}+2\right)  z^{2}+\left(  3c_{1}%
^{2}+4c_{1}+1\right)  z-c_{2}:=P_{2}\left(  z\right)  .
\]

Having obtained $P_{2},$ we use system $\left(  \ref{s2}\right)  $ and let
$f=P_{2}=-\phi,F=\Phi^{+}.$ Solving the equation
\[
Df\cdot F=-f^{+}F^{-}+fF.
\]
we get
\begin{align*}
F\left(  z\right)   =~&z^{6}+6c_{1}z^{5}+\frac{5}{3}\left(  9c_{1}
^{2}-1\right)  z^{4}+\frac{1}{9}\left(  135c_{1}^{3}-90c_{1}^{2}
-105c_{1}-45c_{2}-2\right)  z^{3}\\
&  -\frac{1}{9}\left(  270c_{1}^{3}+225c_{2}^{2}+135c_{1}c_{2}+6c_{1}%
-4\right)  z^{2}\\
&  +\left(  \frac{1}{9}\left(  135c_{1}^{3}+180c_{1}^{2}+270c_{1}c_{2}
+57c_{1}+135c_{2}+4\right)  +c_{3}\right)  z\\
&  +\frac{1}{9}\left(  90c_{1}^{2}c_{2}+45c_{1}c_{2}-45c_{2}^{2}%
-4c_{2}\right)  +c_{3}c_{1}\\
:=~&P_{3}\left(  z\right)  .
\end{align*}
Repeating this procedure, we get polynomial solutions $P_{j},$ $j=1,2,...,$%
with degree $\frac{j\left(  j+1\right)  }{2}.$

With this new formulation, we find the following degree 15 polynomial with odd
symmetry:
\begin{align*}
P_{5}\left(  z\right)   &  =z^{15}-\frac{35}{3}z^{13}+\frac{427}{9}
z^{11}-\frac{6835}{81}z^{9}+\frac{5584}{81}z^{7}\\
&  -\frac{5608}{243}z^{5}+\frac{29744}{10935}z^{3}+\frac{15376}{32805}z.
\end{align*}

We remark that the solutions with degree $k\left(  k+1\right)  /2$ also have
an explicit Wronskian form%
\[
W\left(  q_{1},...,q_{k}\right)  ,
\]
where $q_{k},$ $k\geq1,$ is a polynomial of degree $2k-1$ with
\[
q_{k}\left(  z\right)  =z^{2k-1}+\sum\limits_{j=0}^{k-1}\left(  \alpha
_{j}z^{2j}\right)  .
\]
This can be used to generate solutions in a more efficient way, since in each
step we already have the information on $q_{1},...,q_{k-1}$ and only need to
find $q_{k}$. However, the precise expressions will be more complicated then
that of the generalized Adler-Moser polynomials obtained in \cite{LiuW2}.

One can check using a numerical computation that the balancing configuration
provided by the roots of the polynomials $P_{j}$ are nondegenerated, in the
space of symmetric configurations, at least when $j$ is not large. Therefore,
we can construct family of Riemann type minimal surfaces, stated in Theorem
\ref{main} by perturbing method(either using Weierstrass representation as
done by Traizet, or using PDE methods). This step is more or less standard and
we omit the details. Note that proving nondegeneracy is always a delicate
issue. The polynomials $P$ satisfy the balancing condition always come in
families. Hence the nondegeneracy is only true for the symmetric one in the
space of symmetric configuration. Here it is also possible to prove the
nondegeneracy using Backlund transformation, in a similar spirit as that of
\cite{LiuW}, for the Adler-Moser polynomials.

A modified version of the above polynomial method can be used to find balanced
configurations of soliton type. Let us explain this in more details.

The previous discussion tells us that we can not find balanced singly periodic
configuration such that between each two planes $L_{k}$ and $L_{k+1}$ there
are two catenoids of the same size(say, normalized such that the logrithmic
growth is equal to $1$).

However, as we will show below, there are balanced configuration such that
between $L_{0}$ and $L_{1},$ we put a catenoid with logrithmic growth rate
equals $2$(centered at origin)$,$ and between other planes, there are two
catenoids with logrithmic growth $1.$ Suppose their position are determined by
$p_{k},q_{k}.$

We define, for $k\geq1,$
\[
P_{k}\left(  z\right)  =\left(  z-p_{k}\right)  \left(  z-q_{k}\right)
:=z^{2}+a_{k}z+b_{k}.
\]
The balancing condition implies the following identity: For $k\geq2,$ at
$p_{k}$ and $q_{k}$,
\begin{equation}
\frac{P_{k}^{\prime\prime}}{P_{k}^{\prime}}-\frac{P_{k-1}^{\prime}}{P_{k-1}%
}-\frac{P_{k+1}^{\prime}}{P_{k+1}}=0. \label{res}%
\end{equation}
This can also be view as a residue condition for an identity similar to
$\left(  \ref{infisum}\right)  .$ Since $P_{k}$ is a degree two polynomial,
the equation $\left(  \ref{res}\right)  $ has the following simple form:
\[
\frac{2}{2p_{k}+a_{k}}-\frac{2p_{k}+a_{k-1}}{p_{k}^{2}+a_{k-1}p_{k}+b_{k-1}%
}-\frac{2p_{k}+a_{k+1}}{p_{k}^{2}+a_{k+1}p_{k}+b_{k+1}}=0.
\]
This can be further simplified by repeated use of the fact that $p_{k}%
^{2}+a_{k}p_{k}+b_{k}=0.$ As a consequence
\[
2-\frac{\left(  2p_{k}+a_{k-1}\right)  \left(  2p_{k}+a_{k}\right)  }{\left(
a_{k-1}-a_{k}\right)  p_{k}+b_{k-1}-b_{k}}-\frac{\left(  2p_{k}+a_{k+1}%
\right)  \left(  2p_{k}+a_{k}\right)  }{\left(  a_{k+1}-a_{k}\right)
p_{k}+b_{k+1}-b_{k}}=0.
\]
Therefore,%
\[
2-\frac{2\left(  -a_{k}+a_{k-1}\right)  p_{k}+a_{k}a_{k-1}-4b_{k}}{\left(
a_{k-1}-a_{k}\right)  p_{k}+b_{k-1}-b_{k}}-\frac{2\left(  -a_{k}%
+a_{k+1}\right)  p_{k}+a_{k}a_{k+1}-4b_{k}}{\left(  a_{k+1}-a_{k}\right)
p_{k}+b_{k+1}-b_{k}}=0.
\]
We finally get an equation of the form
\[
A_{k}p_{k}+B_{k}=0,
\]
where
\[
A_{k}=-\left(  a_{k-1}-a_{k}\right)  \left(  a_{k}a_{k+1}-4b_{k}\right)
-\left(  a_{k+1}-a_{k}\right)  \left(  a_{k}a_{k-1}-4b_{k}\right)  +2\left(
a_{k}-a_{k-1}\right)  \left(  a_{k}-a_{k+1}\right)  a_{k},
\]%
\begin{align*}
B_{k}  =~&2\left(  a_{k}-a_{k-1}\right)  \left(  a_{k}-a_{k+1}\right)
b_{k}+2\left(  b_{k-1}-b_{k}\right)  \left(  b_{k+1}-b_{k}\right) \\
&  -\left(  a_{k}a_{k-1}-4b_{k}\right)  \left(  b_{k+1}-b_{k}\right)  -\left(
a_{k}a_{k+1}-4b_{k}\right)  \left(  b_{k-1}-b_{k}\right)  .
\end{align*}
Since we assume $p_{k}\neq q_{k},$ we should have: For $k\geq2,$%
\begin{equation}
\begin{cases}
A_{k}=0,\\
B_{k}=0.
\end{cases}  \label{rec}%
\end{equation}
This is a recurrence relation which can used to determine $\left(  a_{k}%
,b_{k}\right)  ,k\geq3,$ provided that we know the initial condition $\left(
a_{1},b_{1}\right)  $ and $\left(  a_{2},b_{2}\right)  .$ Indeed, one can show
that $a_{k}$ is linear in $k.$ This method can be generalized to polynomials
with higher degrees.

To see how to use the above formulation to get a balanced configuration. Let
us assume
\[
p_{1}=1,q_{1}=\alpha,
\]
where $\alpha\neq0$ and $\alpha\neq1.$ Then the balancing condition for
$p_{1}$ and $q_{1}$ reads as%
\[
\begin{cases}
\frac{2}{p_{1}-q_{1}}=\frac{2}{p_{1}}+\frac{1}{p_{1}-p_{2}}+\frac{1}%
{p_{1}-q_{2}},\\
\frac{2}{q_{1}-p_{1}}=\frac{2}{q_{1}}+\frac{1}{q_{1}-p_{2}}+\frac{1}%
{q_{1}-q_{2}}.
\end{cases}
\]
Solve this system, we get
\[
p_{2}=\alpha+1+\sqrt{\alpha^{2}-\alpha+1},q_{2}=\alpha+1-\sqrt{\alpha
^{2}-\alpha+1}.
\]
Then we can use $\left(  \ref{rec}\right)  $ to get the sequence $\left(
a_{k},b_{k}\right)  $ and $\left(  p_{k},q_{k}\right)  ,k\geq2.$ For
$k\leq-1,$ we take $q_{k}=p_{-k},$ $p_{k}=q_{-k}.$ The resulted balanced
configuration gives a minimal surface of 2-soliton type. They correspond to
the $2$-soliton solutions of the Toda lattice. Taking into account of the fact
that Toda lattice also have $n$-soliton solutions, we should also be able to
construct these type of minimal surfaces using the method proposed in this paper.

\bigskip

\begin{figure}[ptb]
\centering
{ \includegraphics[
		height=0.4\textheight,
		width=0.7\textheight
		\caption{The location of catenoids for a 2-soliton solution.}
		]{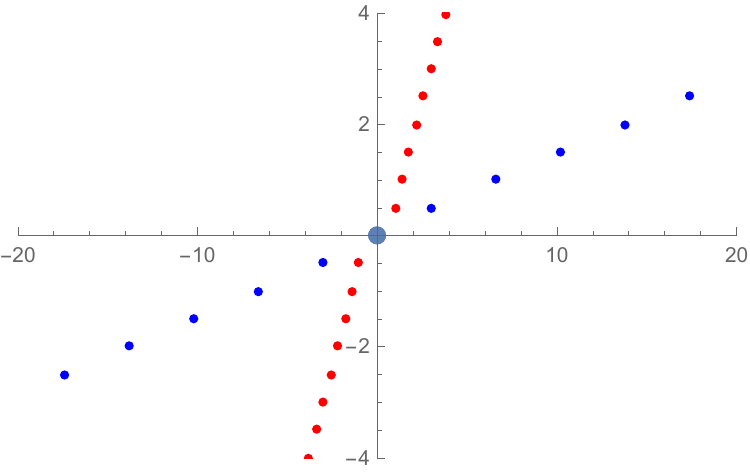} }\end{figure}


\begin{thebibliography}{99}                                                                                               %


\bibitem {LiuW2}W. Ao, Y. Huang, Y. Liu, J. Wei, Generalized Adler--Moser
Polynomials and Multiple Vortex Rings for the Gross--Pitaevskii Equation, SIAM
J. Math. Anal. 53 (2021) 6959--6992. https://doi.org/10/gnzkrt.

\bibitem {Bambusi}D. Bambusi, T. Kappeler, T. Paul, From Toda to KdV,
Nonlinearity 28 (2015) 2461--2496. https://doi.org/10.1088/0951-7715/28/7/2461.

\bibitem {Ca}M. Callahan, D. Hoffman, W.H. Meeks, Embedded minimal surfaces
with an infinite number of ends, Invent Math 96 (1989) 459--505.

\bibitem {CM}O. Chodosh, C. Mantoulidis, Minimal surfaces and the Allen-Cahn
equation on 3-manifolds: index, multiplicity, and curvature estimates, Annals
of Mathematics 191 (2020) 213. https://doi.org/10.4007/annals.2020.191.1.4.

\bibitem {Calla}M. Callahan, D. Hoffman, W.H. Meeks, The structure of
singly-periodic minimal surfaces, Invent Math 99 (1990) 455--481. https://doi.org/10.1007/BF01234428.

\bibitem {Manuel}M. Del Pino, M. Kowalczyk, F. Pacard, J. Wei, Multiple-end
solutions to the Allen--Cahn equation in R 2, Journal of Functional Analysis
258 (2010) 458--503. https://doi.org/10.1016/j.jfa.2009.04.020.

\bibitem {K-D-W}M. Del Pino, M. Kowalczyk, J. Wei, On De Giorgi's conjecture
in dimension $N\geq9$, Ann. of Math. (2)174(2011), no.3, 1485--1569.

\bibitem {KDW2}M. Del Pino, M. Kowalczyk, J. Wei, Entire solutions of the
Allen-Cahn equation and complete embedded minimal surfaces of finite total
curvature in $\mathbb{R}^{3}$, J. Differential Geom. 93(2013), no.1, 67--131.

\bibitem {Flaschka1}H. Flaschka, The Toda lattice. I. Existence of integrals,
Phys. Rev. B (3)9(1974), 1924--1925.

\bibitem {Flaschka2}H. Flaschka, On the Toda lattice. II. Inverse-scattering
solution, Progr. Theoret. Phys. 51(1974), 703--716.

\bibitem {FM}C. Frohman, W. Meeks, The topological classification of minimal
surfaces in $\mathbb{R}^{3},$ Ann. Math. 167 (2008) 681--700. https://doi.org/10.4007/annals.2008.167.681.

\bibitem {Gui}C. Gui, Y. Liu, J. Wei, Two-end solutions to the Allen--Cahn
equation in $\mathbb{R}^{3}$, Advances in Mathematics 320 (2017) 926--992. https://doi.org/10/gcj8kb.

\bibitem {Pacard}L. Hauswirth, F. Pacard, Higher genus Riemann minimal
surfaces, Invent. Math. 169 (2007) 569--620. https://doi.org/10.1007/s00222-007-0056-z.

\bibitem {Hirota}R. Hirota, Toda Molecule Equations, in: Algebraic Analysis,
Elsevier, 1988: pp. 203--216. https://doi.org/10.1016/B978-0-12-400465-8.50024-9.

\bibitem {HS}R. Hirota, J. Satsuma, A Variety of Nonlinear Network Equations
Generated from the Backlund Transformation for the Toda Lattice, Prog. Theor.
Phys. Suppl. 59 (1976) 64--100. https://doi.org/10.1143/PTPS.59.64.

\bibitem {Pa2}S. Kaabachi, F. Pacard, Riemann minimal surfaces in higher
dimensions, Journal of the Inst. of Math. Jussieu 6 (2007) 613. https://doi.org/10.1017/S1474748007000060.

\bibitem {Kons}B. Kostant, The solution to a generalized Toda lattice and
representation theory, Advances in Mathematics 34 (1979) 195--338. https://doi.org/10.1016/0001-8708(79)90057-4.

\bibitem {Kowal}M. Kowalczyk, Y. Liu, J. Wei, Singly Periodic Solutions of the
Allen-Cahn Equation and the Toda Lattice, Communications in Partial
Differential Equations 40 (2015) 329--356. https://doi.org/10.1080/03605302.2014.947379.

\bibitem {LiuW}Y. Liu, J. Wei, Multivortex Traveling Waves for the
Gross--Pitaevskii Equation and the Adler--Moser Polynomials, SIAM J. Math.
Anal. 52 (2020) 3546--3579. https://doi.org/10.1137/18m119940x.

\bibitem {Liu}Y. Liu, J. Wei, W. Yang, Uniqueness of lump solution to the KP-I
equation, Proceedings of London Math Soc 129 (2024) e12619. https://doi.org/10.1112/plms.12619.

\bibitem {Lopez}F. Lopez, M. Ritore, Fusheng Wei, A characterization of
Riemann's minimal surfaces, J. Differential Geom. 47(1997), no.2, 376--397.

\bibitem {Meeks}W. H. Meeks III, J. Perez, A. Ros, Uniqueness of the Riemann
minimal examples, Inventiones Mathematicae 133 (1998) 107--132.

\bibitem {Mora}F. Morabito, M. Traizet, Non-periodic Riemann examples with
handles, Advances in Mathematics 229 (2012) 26--53. https://doi.org/10.1016/j.aim.2011.09.001.

\bibitem {Moser}J. Moser, Finitely many mass points on the line under the
influence of an exponential potential -- an integrable system, in: J. Moser
(Ed.), Dynamical Systems, Theory and Applications, Springer Berlin Heidelberg,
Berlin, Heidelberg, 1975: pp. 467--497. https://doi.org/10.1007/3-540-07171-7\_12.

\bibitem {Olsh}Olshanetsky, M. A.; Perelomov, A. M. Explicit solutions of
classical generalized Toda models.Invent. Math.54(1979), no.3, 261--269.

\bibitem {Popowicz}Z. Popowicz, The Toda lattice and Kadomtsev-Petviashvili
equations, J. Phys. A: Math. Gen. 22 (1989) 5007--5015. https://doi.org/10.1088/0305-4470/22/22/025.

\bibitem {Rogers}C. Rogers, W.F. Shadwick, Backlund transformations and their
applications, Academic Press, New York, 1982.

\bibitem {Toda}M. Toda, Theory of Nonlinear Lattices, Springer Berlin
Heidelberg, Berlin, Heidelberg, 1989. https://doi.org/10.1007/978-3-642-83219-2

\bibitem {T1}M. Traizet, An Embedded Minimal Surface with no Symmetries, J.
Differential Geom. 60 (2002). https://doi.org/2021050905245500504.

\bibitem {T2}M. Traizet, Adding handles to Riemann's minimal surfacess,
Journal of the Institute of Mathematics of Jussieu (2002) 1(1), 145--174.

\bibitem {V1}J. Villarroel, M. J. Ablowitz, On the method of solution to the
2+ 1 Toda equation, Physics Letters A 163 (1992) 293--298.

\bibitem {V2}J. Villarroel, M. J. Ablowitz, On the inverse scattering
transform of the 2 + 1 Toda equation, Physica D 65 (1993), 48--70.

\bibitem {V3}J. Villarroel, M. J. Ablowitz, Solutions to the 2+1 Toda
equation, J. Phys. A: Math. Gen. 27 (1994) 931--941. https://doi.org/10.1088/0305-4470/27/3/032.
\end{thebibliography}
\end{document}